\documentclass[12 pt]{article}
\usepackage{amsmath, amsfonts, amssymb}
\usepackage{graphicx}
\usepackage{indentfirst}
\usepackage{threeparttable}
\usepackage{url}
\begin{document}
 \title{\LARGE \bf Three Models of the Gravitational Potential of the Milky Way}
\bigskip
 \bigskip
 \author{\bf V.~V.~Bobylev\thanks{E-mail: bob-v-vzz@rambler.ru},
A.~T.~Bajkova 
and
A.~A.~Smirnov
}
\date{\it  \small  $^1$ Pulkovo Astronomical Observatory, \\
St.-Petersburg 196140, Russia}

\maketitle

\begin{abstract}
The parameters of an axisymmetric model for the gravitational potential of the Galaxy have been
refined. The basic curve of the Galaxy's rotation in a distance interval of $R:0-190$ kpc was constructed using
the velocities of masers, classical Cepheids, Red Clump stars, Blue Horizontal Branch stars, halo stars, globular
clusters, and dwarf satellite galaxies of the Milky Way. The rotation curve was selected in such a way that
there would be no dominant burst of circular velocities in the central ($R<2$ kpc) region of the Galaxy. As a
result, we constructed two two-component models of the galactic potential, which include contributions
from the disk and the halo of invisible matter, as well as a three-component model with a small-mass bulge
added in advance. These models can be useful in studying the long-term orbital evolution of stars and open
and globular star clusters in the central ($R<4$ kpc) region of the Galaxy. The constructed models were tested
for self-consistency by comparing their rotation curves with a set of model curves generated with the Illustris
TNG50 software package.
\end{abstract}


 \newpage
 \section{INTRODUCTION}

A model of the gravitational potential of the Galaxy
is of great importance for studying its structure and
dynamics. One of the main data sources for constructing
such a model is the circular rotation velocities $V_{circ}$
of objects located at different distances $R$ from the
rotation axis of the Galaxy. The dependence of $V_{circ}$ on
distance $R$ is called the galactic rotation curve.

At present, various models of the gravitational
potential of the Galaxy have been proposed. Mostly,
these are axisymmetric multicomponent models [1--5]. There are also nonaxisymmetric models, which
take into account the influence of the bar and the spiral
density wave (e.g., [6]), as well as models, in which
the circular rotation velocities $V_{circ}$ of stars change over
time due to the influence of the Large Magellanic
Cloud; the latter has a noticeable impact on the estimate
of the Galaxy mass [7, 8].

In papers [3, 9, 10] six axisymmetric three-component
(a bulge, a disk, and a halo) models of the gravitational
potential of the Galaxy were considered. To
refine the parameters of these models, observational
data covering a range of distances $R$ from 0 to
$\sim$200 kpc were used. At distances $R<20$~kpc, the
emphasis was given to the data on the radial velocities
of hydrogen clouds at tangential points and the data on
130 masers, for which the trigonometric parallaxes are
available; while for greater distances, the averaged
rotation velocities of the Galaxy from a review [1] were
used.

On the rotation curve of the Galaxy, there is a well-known
burst in the central region at $R<2$ kpc. This
burst was identified exclusively from the data on neutral
hydrogen clouds at tangential points (e.g., [11,
12]). Recently, in a number of studies, it has been
shown that the burst on the rotation curve of the Galaxy
may appear due to errors in the method of determining
the circular velocities from tangential points in
a nonaxisymmetric disk within a region of 3--4 kpc
([13], Fig. 9).

On the other hand, according to some authors, the
burst of circular velocities of stars in the central region
of the Galaxy depends on the mass of the bulge (see,
e.g., [14]) and is also associated with the influence and
orientation of the central bar (see, e.g., [15]). It is
known that the bar rotates with a constant angular
velocity, the value of which exceeds the angular rotation
velocity of the Galaxy in the vicinity of the Sun.
Consequently, the orientation of the bar relative to the
Sun-Galactic center direction changes over time.
According to the results of modeling [15], when the
orientation of the bar changes, the peak of velocities
can be replaced by a deep minimum. In the end, many
authors now prefer to use models of the gravitational
potential of the Galaxy with the rotation curve exhibiting
no strong oscillations in the central region (see,
e.g., [6, 16--20]).

The aim of this work is to construct such models of
the gravitational potential of the Galaxy, in which the
rotation curve does not have a significant peak of
velocities in the center. The need for such a model
arises when studying the long-term orbital evolution of
stars and open and globular star clusters in the central
($R<4$~kpc) region of the Galaxy, where there is a
strong influence of the bar.

\section{DATA}

In this paper, to construct the rotation curve of the
Galaxy, we use the positions and rotation velocities
(around the center of the Galaxy) of various galactic
objects, data on which we took from literary sources.
These are (a) maser sources and radio stars with measured
trigonometric parallaxes, (b) classical Cepheids,
(c) Red Clump giants, Blue Horizontal Branch giants,
and distant halo stars, as well as (d) globular clusters
and dwarf satellite galaxies of the Milky Way.

\subsection{A Sample of Masers and Radio Stars}

The use of the very-long-baseline interferometry
(VLBI) method to measure the trigonometric parallaxes
of galactic masers made them first-class objects
for studying the Galaxy. Of greatest interest are maser
sources associated with young stars and protostars
located in active star-forming regions. A sample of
approximately 200 sources of this kind is presented
and analyzed in a paper [21]. According to that study,
random errors in VLBI measurements of the maser
trigonometric parallaxes are less than 10\% on average.

After the appearance of a fundamental study [21],
several more radio astronomy papers were published
on the measurement of trigonometric parallaxes of
masers (see, e.g., [22--25]). We also considered highprecision
VLBI measurements of parallaxes of young
radio stars in the continuum (e.g., [26]). As a result,
our sample contains 265 measurements of trigonometric
parallaxes of radio objects, the random measurement
errors of which are less than 30\%. In this
paper, from this sample, we use 196 objects, for which
parallax errors are less than 15\% and which are located
further than 4 kpc from the Galactic center. Furthermore,
masers, for which the radial velocity errors
exceed 20 km/s, were not considered. We divided the
masers into nine intervals (more than five sources in
each) by distance $R$, according to which the average
values of circular velocities and their errors were calculated
using the 3$\sigma$ criterion to reject leaps. Note that
no leaps were observed in this sample, since we had
previously discarded the data with large errors in the
measurement of parallaxes and radial velocities.

\subsection{A Sample of Classical Cepheids}

For a large number of classical Cepheids, high-precision
heliocentric distances were determined in [27]
by using the period-luminosity relationship. Observations
of these Cepheids were carried out within the
framework of the Optical Gravitational Lensing
Experiment (OGLE) program [28, 29]. Currently, the
updated list of observations already includes more
than 3500 stars [30]. In papers [27, 31] it was shown
that, in the catalogue, random errors in determining
the distances to Cepheids by internal convergence are
less than 5\%, while by external convergence (from
comparison with trigonometric parallaxes of Cepheids
from the Gaia DR3 catalogue [32]), less than 10\%.

Bobylev and Bajkova [33] obtained a sample of
about 2000 stars by identifying Cepheids from the list
[27] with the Gaia EDR3 catalogue [34], from which
the proper motions of Cepheids were copied. We took
the radial velocities of 773 Cepheids from a paper [35].
We use this sample in the present paper. Note that in
[35] the proper motions of Cepheids from the Gaia
DR2 catalogue [36] were used. We divided the Cepheids
by distance $R$ into 15 intervals (with more than four
stars in each). In each of the intervals, the average values
of rotation velocities around the galactic center
and their errors were calculated with disregarding
jumps according to the 3$\sigma$ criterion if necessary.

Two stars, V371~Per and V800~Aql, stand out from
the entire list: their rotation velocities significantly
deviate from the average value in the corresponding
distance range. For V371 Per, the deviation of the rotation
velocity $V_{circ}$ from the average in the range is
43.1 km/s (the 3.6$\sigma$ level for the corresponding dispersion);
since this star falls into the distance range that
contains 125 stars, removing it from the sample has no
noticeable effect on the average value calculated. So,
excluding it from the data set changes the velocity
value $V_{circ}$ by 0.3 km/s. For this star, the radial and tangential
velocities, $V_R=1.8\pm5.4$~km/s and $V_{circ}=268.6\pm3.3 $~km/s, were calculated with very small
errors.

For V800~Aql, the deviation of the velocity $V_{circ}$
from the average in the range is $-54.2 $~km/s (the
3.2$\sigma$ level for the corresponding dispersion); in this
range, there are 19 stars, so its influence on the calculation
of the average is not critical either. Removing it
from the data set changes the rotation velocity value
$V_{circ}$ by 3 km/s. For this star, the radial and tangential
velocities are calculated with fairly large errors: $V_R=37.2\pm18.0$~km/s and $V_{circ}=168.7\pm17.7$~km/s. As a
result, we decided not to discard V371~Per and
V800~Aql.

\subsection{A Sample of Red Giants}

A sample of more than 250000 high-luminosity red
giants was studied in the paper of Zhou et al. [37],
where this sample is designated as Luminous Red
Giant Branch (LRGB). These are stars belonging to
the asymptotic branch of giants, and they are above
the Red Clump giants on the Hertzsprung-Russell
diagram. To construct the rotation curve, Zhou et al.
[37] considered 54 000 stars of this kind and their photometric
distances. They took the necessary photometric
and spectral (radial velocities) data from the
APOGEE [38] and LAMOST [39, 40] surveys.
According to Zhou et al. [37], random errors in determining
the distances to the selected stars amounted to
10--15\%.

As for the circular rotation velocities of stars
around the center of the Galaxy $V_{circ} (R)$, Zhou et al.
[37] estimated them indirectly on the basis of the Jeans
equation [41] through velocity dispersions:
 \begin{equation}
\begin{array}{lll}
 \renewcommand{\arraystretch}{3.2}
 \displaystyle
  V^2_{circ} (R)=\langle V^2_{\theta}\rangle  - \\
 \displaystyle
    \qquad  -\langle V^2_{R}\rangle  \Biggl(1+ \frac{\partial\ln \rho}{\partial\ln R}+\frac{\partial \ln \langle V^2_{R}\rangle}{\partial\ln R} \Biggr),
  \label{Jeans-1}
 \end{array}
 \end{equation}
where the equations are written in a cylindrical coordinate
system $(R,\theta,z)$; $\rho (R,z)$ is the stellar density;
$V_R$ and $V_{\theta}$~ are the corresponding velocities of stars; and
$\langle V^2_{R}\rangle$ and $\langle V^2_{\theta}\rangle$~ are the averages of the squares of the
corresponding velocities of stars. The second term on
the right-hand side of relationship (1) is called the
asymmetric drift.

This method is model-dependent, since it is necessary
to know well the distribution law for the stellar
density in dependence on $R$ and $z$ and the behavior of
the velocity dispersions of the analyzed stars in dependence
on $R$. Usually, all these characteristics are well-known only in the immediate vicinity of the Sun
(near $R_0$).

It is worth noting the papers [42, 43], in which the
rotation curve of the Galaxy was constructed on the
basis of relationships (1) using the Gaia DR3 catalogue
data. In [42], the velocities of stars, belonging to
the clump of red giants located at distances in a range
of $R:5-14$~kpc, were analyzed. In [43], the rotation
curve of the Galaxy was constructed using the Gaia
DR3 data on stars, for which the radial velocities are
available and which are in an interval of $R<21$~kpc. It
is important that the rotation curves of the Galaxy
constructed in [43] well agree with the curves constructed
directly (see relationship (3) below).

\subsection{Data from the Paper of Bhattacharjee et al. [1]}

Bhattacharjee et al. [1] constructed a rotation curve
of the Galaxy in a range of galactocentric distances of
0-200 kpc on the basis of a variety of kinematic data.
In this study, at distances $R$ exceeding 25 kpc, the
radial velocities of the following objects of the thick
disk and the halo were used: 1457 Blue Horizontal
Branch giants, 2227 K-giants, 16 globular clusters,
28 distant halo stars, and 21 dwarf satellite galaxies of
the Milky Way. In [1] the rotation curve of the Galaxy
was constructed with the values $R_0=8.3$~kpc and $V_0=244$~
km/s. The circular rotation velocities $V_{circ} (R)$
of distant objects located at distances exceeding 25 kpc
were estimated with relationship (1).

 {\begin{table}[t]                                    
 \caption[]
 {\small\baselineskip=1.0ex
 Estimates of $R_0$ and $V_0$.  }
 \label{t-1}
 \small{\begin{center}\begin{tabular}{|l|c|c|c|c|}\hline
      $R_0$,~kpc      & $V_0$,~km/s   & $\partial V/\partial R$  & $n_\star$   & Ref \\
           &    & km/s/kpc  &    &  \\\hline
  $8.15\pm0.15$     & $236\pm7$      &                           & $\sim$200 masers            & [17]\\
  $8.122\pm0.031$ & $233.6\pm2.8$ & $-1.34\pm0.21$  &           773 Cepheids        & [31]\\
  $8.122\pm0.031$ & $229.0\pm0.2$ & $-1.7  \pm0.1 $  & $\sim$20000 red giants & [40]\\
  $8.122\pm0.031$ & $234.0\pm1.4$ & $-1.84 \pm0.07$ & $\sim$54000 giants of the branch  & [33]\\
  $8.122\pm0.031$ & $235.2\pm0.8$ & $-1.33\pm0.1 $  & $\sim$3500 Cepheids            & [4]\\
  $ 8.1  \pm0.1$     & $236.3\pm3.3$ &                          &              770 Cepheids      & [29]\\
\hline
 \end{tabular}\end{center}
 }
  \end{table}}

\subsection{Rotation Velocities of Stars
Around the Galactic Center}

For each star, the observations give the radial
velocity $V_r,$ directed along the line of sight and two projections
of the tangential velocity, $V_l=4.74r\mu_l\cos b$
and $V_b=4.74r\mu_b,$, directed along the galactic longitude $l$
and latitude $b$, respectively. The coefficient 4.74
is a ratio of the number of kilometers in an astronomical
unit to the number of seconds in a tropical year.
Through the components $V_r, V_l, V_b$ we can calculate
the velocities $U,V,W$ directed along the rectangular
galactic coordinate axes:
 \begin{equation}
 \begin{array}{lll}
 U=V_r\cos l\cos b-V_l\sin l-V_b\cos l\sin b,\\
 V=V_r\sin l\cos b+V_l\cos l-V_b\sin l\sin b,\\
 W=V_r\sin b                +V_b\cos b,
 \label{UVW}
 \end{array}
 \end{equation}
where the velocity $U$ is directed from the Sun parallel
to the direction to the Galaxy center, or, more precisely,
to the rotation axis of the Galaxy; $V$ is oriented
in the direction of rotation of the Galaxy; and $W$
points towards the north galactic pole. The velocities
calculated with relationships (2) are given relative to
the Sun. To obtain the velocities relative to the local
standard of rest (LSR), it is necessary to take into
account the motion of the Sun relative to the LSR
with the velocity components $(U,V,W)_\odot$. Currently,
the values of the components of the peculiar velocity
of the Sun relative to the LSR $(U,V,W)_\odot=(11.1,12.24,7.25)$~
km/s, which were determined in
[44], are widely used.

To correctly calculate the circular velocities, it is
necessary to take into account the correction for
asymmetric drift (1). The value of this correction
depends on the age of a sample of stars. For the youngest
stars (e.g., masers, OB stars, or young Cepheids)
this correction is close to zero, while for the oldest
galactic objects (e.g., halo stars or globular clusters) its
value is close to 200 km/s. In this work we did not take
into account the corrections for asymmetric drift for
masers and Cepheids. However, Zhou et al. [37] and
Bhattacharjee et al. [1] considered them for red giants
and a variety of samples of stars, respectively.

The value of the rotation velocity of a star around
the galactic center $V_{\rm circ} (R)$ (its direction coincides with
the direction of the Galaxy rotation), can be found
from the expression
 \begin{equation}
 \begin{array}{lll}
 \renewcommand{\arraystretch}{2.0}
  V_{\rm circ}= U\sin \theta+(V_0+V)\cos \theta,
 \label{Vcirc}
 \end{array}
 \end{equation}
where the position angle satisfies the relationship
$\tan\theta=y/(R_0-x)$, $x,y,z$~ are the rectangular heliocentric
coordinates of the star: the $x$-axis is directed
from the Sun to the center of the Galaxy, the $y$-axis, in
the direction of the Galaxy rotation, and the $z$-axis, to
the north galactic pole (the velocities $U,V,W$) are
directed along the corresponding axes $x,y,z$);
and $V_0$ is the linear circular rotation velocity of the
Galaxy at a distance $R_0,$ which is the distance from the
Sun to the rotation axis of the Galaxy.

As can be seen from relationship (3), to calculate
$V_{\rm circ}$, it is necessary to set specific values for two quantities,
$R_0$ and $V_0$. Table 1 provides a brief summary of
the $R_0$ and $V_0$ estimates obtained for various objects.
Note that Bobylev and Bajkova [46] found the value $R_0=8.1\pm0.1$~
kpc from statistical analysis of a large
number of individual estimates. As a result, in this
paper, we accept $R_0=8.1$~kpc and $V_0=236$~km/s.
Consequently, we reduce the velocities of objects from
[1] to the values $R_0=8.1$~kpc and $V_0=236$~ km/s by
appropriate adjusting.

\section{METHOD}
\subsection{Model of the Potential of the Galaxy}

In most cases, the axisymmetric potential of the
Galaxy is represented as a sum of three components,
corresponding to a central spherical bulge $Phi_b(r(R,z))$,
a disk $\Phi_d(r(R,z))$, and a massive spherical halo of dark
matter $\Phi_h(r(R,z))$:
 \begin{equation}
 \begin{array}{lll}
 \renewcommand{\arraystretch}{2.8}
  \Phi(R,z)=\Phi_b(r(R,z))+\\
   \qquad+\Phi_d(r(R,z))+\Phi_h(r(R,z)).
 \label{pot}
 \end{array}
 \end{equation}
Here, a cylindrical coordinate system ($R,\theta,z$) with the
origin at the center of the Galaxy is used. In a rectangular
coordinate system $(x,y,z)$ originating from the
center of the Galaxy, the distance to the star is
$r^2=x^2+y^2+z^2=R^2+z^2$. The gravitational potential
is expressed in units of 100 km$^2$/s$^2$, the distances
are in kpc, and the masses are in galactic mass units
$M_{g}=2.325\times 10^7 M_\odot$, corresponding to the gravitational
constant $G=1$.

The expression for the mass density is derived from
the Poisson equation,
\begin{equation}
  \renewcommand{\arraystretch}{1.2}
4\pi G\rho(R,z)=\nabla^2\Phi(R,z) \label{pois1}
\end{equation}
and takes the following form:
 \begin{equation}
 \begin{array}{lll}
 \renewcommand{\arraystretch}{2.8}
 \displaystyle
 \rho(R,z)=\frac{1}{4\pi G}\Bigg{(}\frac{d^2\Phi(R,z)}{dR^2}+\\
 \displaystyle
 \qquad+\frac{1}{R}\frac{d\Phi(R,z)}{dR}+\frac{d^2\Phi(R,z)}{dz^2}\Bigg{)}.
    \label{pois2}
 \end{array}
\end{equation}
The force acting in the direction $z$, perpendicular to
the plane of the Galaxy, is defined as follows:
\begin{equation}
 \renewcommand{\arraystretch}{1.2}
  K_z (z,R)=-\frac{d\Phi(z,R)}{dz}.
 \label{Kz}
\end{equation}
Expressions (6) and (7) are required for further fitting
the parameters of the gravitational potential models
with the restrictions imposed on the local dynamic
density of matter $\rho_0$ and the force $K_z (z,R_0)$ at
$z=1.1$~kpc, which are known from observations [47, 48].

In addition, we will need the following expressions
to calculate

(1) the circular velocities
\begin{equation}
V_{circ}(R)=\sqrt{R\frac{d\Phi(R,0)}{dR}},
\label{V}
\end{equation}

(2) the mass of the Galaxy contained in a sphere of
radius $r$:
\begin{equation}
m(<r)=r^2\frac{d\Phi(r)}{dr},
\label{m}
\end{equation}

(3) the parabolic velocity, or the velocity of a star
escaping from the gravitational field of the Galaxy
\begin{equation}
V_{esc}(R,z)=\sqrt{-2\Phi(R,z)},
\label{Vesc}
\end{equation}

(4) the Oort parameters
\begin{equation}
A=\frac{1}{2}\Omega_0^{'}~ \hbox{\rm and} ~B=\Omega_0+A,
\label{B}
\end{equation}
where $\Omega=V/R$ is the angular rotation velocity of the
Galaxy ($(\Omega_0=V_0/R_0)$), $\Omega_0^{'}$ is the first derivative of the
angular velocity with respect to $R$, and $R_0$ is the distance
from the Sun to the galactic rotation axis; and

(5) the surface density of gravitating matter within
the distance $z_{out}$ from the galactic plane $z=0$:
 \begin{equation}
 \begin{array}{lll}
  \renewcommand{\arraystretch}{4.5}
   \displaystyle
 \Sigma_{out}(z_{out})= 2\int_0^{z_{out}} \rho(R,z)dz=\\
    \displaystyle
  =K_{z}/(2\pi G)+(B^2-A^2)/(2\pi G),
\label{Sigma}
 \end{array}
\end{equation}
In this paper the potentials of a bulge $\Phi_b(r(R,z))$ and
a disk $\Phi_d(r(R,z))$ are presented in the following forms:
 \begin{equation}
 \renewcommand{\arraystretch}{1.2}
  \Phi_b(r)=-\frac{M_b}{(r^2+b_b^2)^{1/2}},
  \label{bulge}
 \end{equation}
 \begin{equation}
 \Phi_d(R,z)=-\frac{M_d}{\{R^2+[a_d+(z^2+b_d^2)^{1/2}]^2\}^{1/2}},
 \label{disk}
\end{equation}
Expression (13) is called the Plummer potential [49],
and expression (14) was first proposed in [50] to model
the disk. Here $M_b, M_d$~ are the masses of the components,
and $b_b, a_d, b_d$ are the scale parameters of
the components expressed in kpc. The contributions
of a bulge and a disk to the circular velocity are,
respectively,
\begin{equation}
V_{circ(b)}^2(R)=\frac{M_b R^2}{(R^2+b_b^2)^{3/2}},
\label{Vc-b}
\end{equation}
\begin{equation}
V_{circ(d)}^2(R)=\frac{M_d R^2}{(R^2+(a_d+b_d)^2)^{3/2}}. 
\label{Vc-d}
\end{equation}
To model the halo potential, we use the expression
 \begin{equation}
  \Phi_h(r)=-\frac{M_h}{r}\ln {\Biggl(1+\frac{r}{a_h}\Biggr)},
 \label{halo-III}
 \end{equation}
proposed in [51], where the contribution to the circular
velocity is
 \begin{equation}
  V^2_{circ(h)}(R)= M_h \biggl[\frac{\ln(1+R/a_h)}{R}-\frac{1}{R+a_h}\biggr].
 \label{Vc-h-III}
 \end{equation}

\subsection{Adjustment of the Parameters}

As follows from [1], the velocities of all objects on
the rotation curve of the Galaxy were calculated with
the values $R_\odot=8.3$~kpc and $V_\odot=244$~km/s. The model
parameters of the potential are determined by fitting
(with the least squares method) to the measured rotation
velocities $V_{circ}$ of galactic objects, where
\begin{equation}
  V^2_{circ}(R)= V^2_{circ(b)}+V^2_{circ(d)}+V^2_{circ(h)}.
 \label{Vc-all}
 \end{equation}
Together with the surface density $\Sigma_{1.1}$, the local
dynamic density of matter $\rho_\odot$, which is a sum of the
densities of the bulge, disk, and invisible matter in the
small solar neighborhood, are the most important
additional constraints in the problem of adjusting the
model parameters of the potential to the data on circular
velocities measured [52]:
\begin{equation}
\rho_0=\rho_b(R_0)+\rho_d(R_0)+\rho_h(R_0),
\label{ro}
\end{equation}
\begin{equation}
\Sigma_{1.1}=
  \int\limits^{1.1\,\hbox {\footnotesize\it kpc}}_{-1.1\,\hbox {\footnotesize\it kpc}}
(\rho_b(R_0,z)+\rho_d(R_0,z)+\rho_h(R_0,z))dz.
\label{Sig}
\end{equation}
The surface density is closely connected with the force $K_z(z,R)$
in accordance with relationship (12). Since
the values of the two most important parameters $\rho_0$
and $K_z/2\pi G$ are known from observations with a sufficiently
high accuracy, we can significantly refine the
parameters of the gravitational potential by introducing
additional restrictions on these two parameters.

When fitting our models to the measurement data,
we use the following target quantities: (1) the local
dynamic density of matter is restricted to $\rho_0=0.1 M_\odot$~pc$^{-3}$,
the value of which is taken according
to [47], and (2) the force acting perpendicular to
the plane of the Galaxy, which is assumed to be
$K_{z=1.1}/2\pi G=72 M_\odot$~pc$^{-2}$ according to the estimate by
Horta et al. [48]. As a result, we used two additional
constraints, and the problem of adjusting the parameters
was reduced to minimizing the following quadratic
functional $F$:
 \begin{equation}
 \begin{array}{lll}
  \renewcommand{\arraystretch}{4.9}
   \displaystyle
\min F=\sum_{i=1}^N (V_{circ}(R_i)-\widetilde{V}_{circ}(R_i))^2+\\
 \displaystyle
+\alpha_1(\rho_\odot-\widetilde{\rho}_\odot)^2+\alpha_2(K_{z=1.1}/2\pi G-\\
  \displaystyle
-\widetilde{K}_{z=1.1}/2\pi G)^2,
\label{F}
 \end{array}
\end{equation}
where $N$ is the number of data; the measurement data
on circular velocity are indicated by a tilde; $R_i$ is the
galactocentric distances of objects; and $\alpha_1, \alpha_2$ are
the weighting coefficients for additional constraints,
which were selected in such a way as to achieve a minimum
in the discrepancy between the data and the
model rotation curve, provided that additional constraints
are met with an accuracy of no worse than 5\%.
The coefficients $\alpha_1$ and $\alpha_2$ can be selected with
accounting for errors in $\rho_0$ and $K_z$, which just provides
these 5\% in this case.

The errors in all parameters given in Table 2 were
determined by the Monte Carlo statistical simulation
method. At each step, the procedure used 1000 independent
realizations of random errors in the measurement
data obeying the normal law with the zero mean
and the standard deviation $\sigma$ known. Note that the
calculations are performed with weights in a form of
$w=1/\sigma^2,$, where $\sigma$ are errors in the velocity values.

 {\begin{table}[t]                                    
 \caption[]
 {\small\baselineskip=1.0ex
 The values of parameters derived with the two-component (a disk and a halo) and three-component (a bulge, a
disk, and a halo) models of the galactic potential, $M_{\rm gal}=2.325\times 10^7 M_\odot$  }
 \label{t:model}
 \begin{center}\begin{tabular}{|c|c|c|c|c|}\hline
  Parameter                 & Model 2-1              & Model 2-2              &  Model 3 \\\hline
 $M_b$($M_{\rm gal}$) &                               &                                &   203$\pm     5$ \\
 $M_d$($M_{\rm gal}$) &  2362$\pm   17$      &   2376$\pm    16$    &  2364$\pm    14$ \\
 $M_h$($M_{\rm gal}$) &  8091$\pm 246$     &   8257$\pm   233$    &  8687$\pm   224$ \\
 $b_b$(kpc)                  &                               &                                &0.2700$\pm0.0002$ \\
 $a_d$(kpc)                  &   4.20$\pm  0.11$    &   4.13$\pm  0.09$    &  4.30$\pm  0.09$ \\
 $b_d$(kpc)                  & 0.2416$\pm0.0005$ & 0.2453$\pm0.0005$ &0.2740$\pm0.0004$ \\
 $a_h$(kpc)                  &   4.74$\pm  0.19$    &   4.87$\pm  0.18$     &  5.30$\pm  0.17$ \\\hline
\end{tabular}\end{center}\end{table}}

\begin{figure}[t]
{ \begin{center}
  \includegraphics[width=0.9\textwidth]{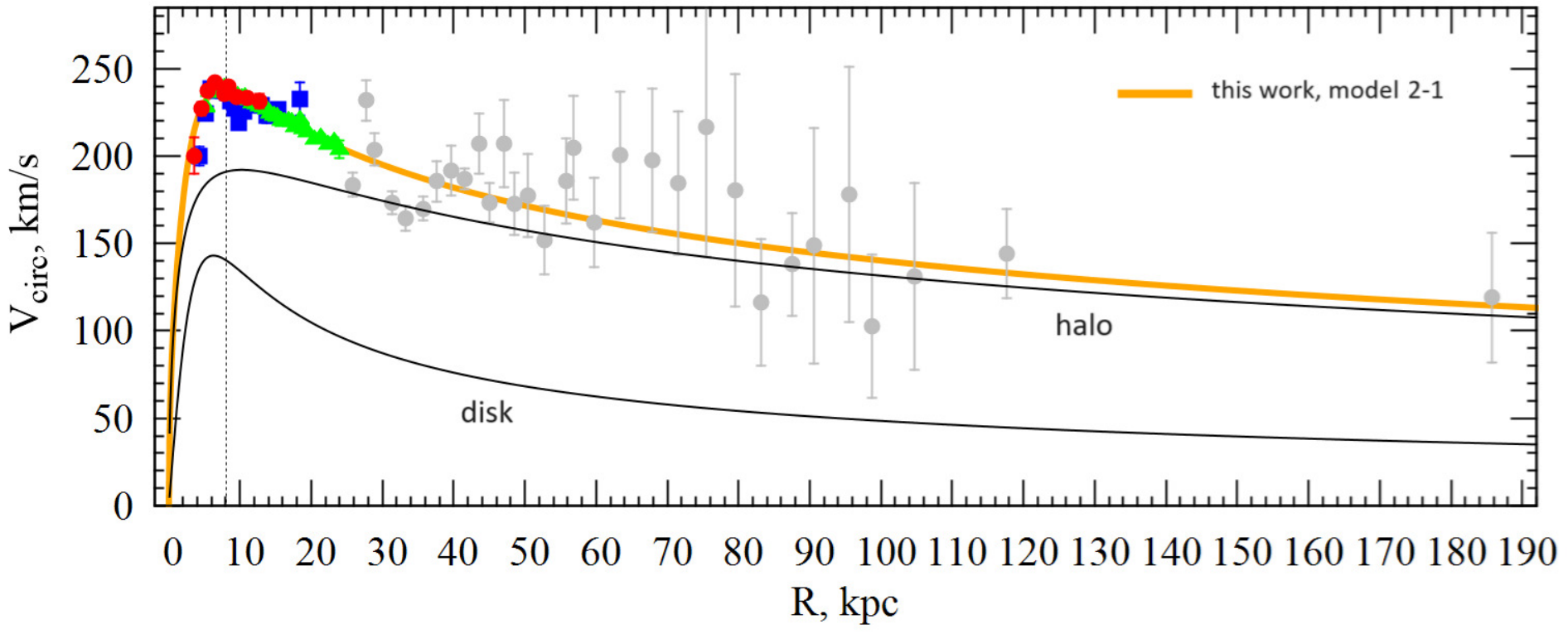}
\caption{The rotation curve of the Galaxy for model 2-1 (orange thick line); the vertical line marks the position of the Sun; the thin
black lines indicate the contributions of the disk and the halo; the velocities of classical Cepheids, masers with measured trigonometric parallaxes, and high-luminosity red giants [37] are shown with blue squares, red circles, and green triangles, respectively, while the gray circles show the velocities according to [1].
}
 \label{f1}
\end{center}}
\end{figure}
\begin{figure}[t]
{ \begin{center}
  \includegraphics[width=0.95\textwidth]{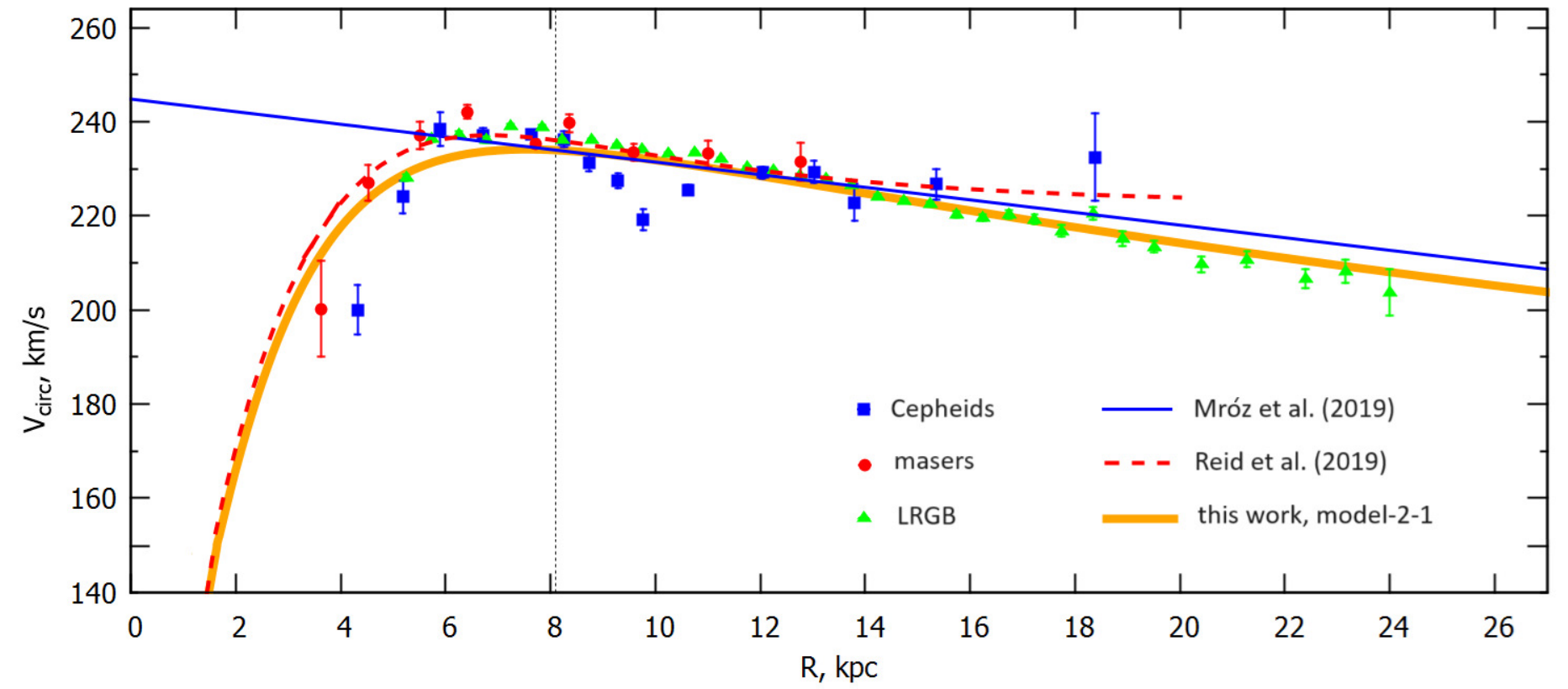}
  \caption{Segments of the rotation curves of the Galaxy in the solar neighborhood, which were found in the studies listed in the legend, in comparison to the rotation curve based on model 2-1; LRGB is the luminous red giant branch.
}
 \label{f2}
\end{center}}
\end{figure}

\section{RESULTS AND DISCUSSION}
 \subsection{Model 2-1}

Since data for $R<3$~kpc are completely lacking, we
decided to construct, first, a two-component model of
the axially symmetric potential of the Galaxy that
includes the contributions of a disk and a halo of invisible
matter. The values of parameters of this two-component
model, which was obtained using all four types
of velocities described above, are given in Table 2.
It is designated as model 2-1. When fitting the data,
the following values of the two target parameters
were obtained: $K_{z=1.1}/2\pi G=73.7~M_\odot$~pc $^{-2}$ and
$\rho_0=0.11 M_\odot$~pc$^{-3}$, which are in good agreement with those
originally sought.

Here, the value of the linear circular rotation velocity
of the Galaxy at the near-solar distance is $V_0=233$~km/s.
With the parameters of model 2-1 we obtain
a number of important estimates. So, the Galaxy mass
contained within a sphere with a radius of 200 kpc is
$M(<200~\hbox {kpc})=(0.68\pm0.18)\times10^{12}M_\odot.$. The escape
velocity at the near-solar distance $R_0=8.1$~kpc is
515 km/s, while it is 200 km/s at a distance of $R=200$~kpc.

For comparison, we note that the estimate $M(<200~\hbox {kpc})=(0.75\pm0.19)\times10^{12}M_\odot$
reported in a paper [9] was obtained with a three-component
potential model (model III) constructed according to
the galactic rotation curve with a peak of velocities in
the central region. As statistical analysis of various
estimates of the Galaxy mass shows, a current value of
$M(<200~\hbox {kpc})$
is close to $1\times10^{12}M_\odot$ [53].

The rotation curves of the Galaxy for model 2-1 are
presented in Fig.~1. Here the velocities of objects from
[1] were reduced to the values $R_0=8.1$~kpc and $V_0=236$~km/s by appropriate adjusting.

The rotation curves of the Galaxy in the solar
neighborhood, which were found in different studies,
are compared in Fig. 2. Let us note a few points
regarding this figure. First, at a given scale, the average
circular velocities of masers, Cepheids, and red giants
are clearly visible in the diagram. Second, the diagram
contains the rotation curve of the Galaxy found from
masers in [21]. Third, there presented a linear trend
revealed by the analysis of the kinematics of 773 classical
Cepheids in [35]. Finally, the rotation curve corresponding
to model 2-1 constructed in this paper is
shown.

We can conclude that, in a region of $R:[6-16]$~kpc,
all three rotation curves are in excellent agreement.
However, when $R>16$~kpc, the discrepancy between
the curves systematically increases. As can be seen,
when $R\sim20$~kpc, the difference between the curves
obtained here and by Reed et al. [21] is about 10 km/s.
This discrepancy is of significant importance in the
spectral analysis of residual rotation velocities, since
the residual velocities of stars is a result of subtracting
the rotation velocities of the Galaxy. For masers, this
is not definitely relevant, since there are still few measurements
of trigonometric parallaxes of distant
masers. However, for Cepheids, this is important,
since kinematic data for quite a large number of distant
Cepheids are available. Perturbations in circular
velocities caused by the galactic spiral density wave are
about 4-6 km/s. Hence, it is necessary to have a "correct"
(smooth, without strong jumps and bends) curve
of the Galaxy rotation over the entire range of distances
$R$, in which the stars under consideration are
present.

\begin{figure}[t]
{ \begin{center}
  \includegraphics[width=0.95\textwidth]{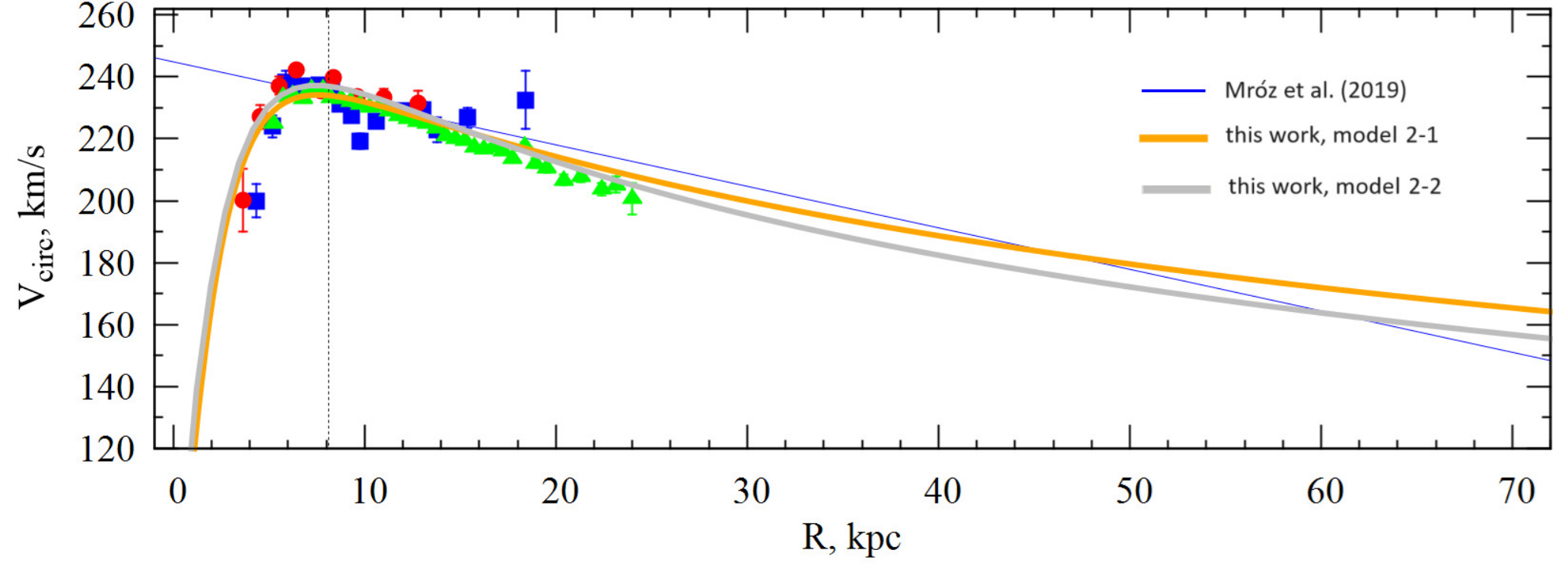}
  \caption{The rotation curves of the Galaxy corresponding to models 2-1 and 2-2.}
 \label{f3}
\end{center}}
\end{figure}

 \subsection{Model 2-2}

Next, we estimated the parameters of the twocomponent
model obtained with only three types of
velocities-for objects that are no further than 25 kpc
from the center of the Galaxy. In other words, the data
[1], containing large errors in the average values
of velocities, were not used. With this approach,
model 2-2 was obtained; and the values of its parameters
are presented in Table 2.

From fitting this model to the data, the following
values were derived for the two target parameters:
$K_{z=1.1}/2\pi G=72.1~M_\odot$~pc$^{-2}$ and $\rho_0=0.11 M_\odot$~pc$^{-3}$.
Here, the Galaxy mass within a sphere of radius
200 kpc is $M(<200~\hbox {kpc})=(0.47\pm0.12)\times10^{12}M_\odot.$
The escape velocity is 477 km/s at the near-solar distance $R_0=8.1$~kpc,
while it is 161 km/s at a distance of $R=200$~kpc.

The rotation curves of the Galaxy in models 2-1
and 2-2 are shown in Fig. 3. As can be seen from the
diagram, the rotation curve of the Galaxy corresponding
to model 2-2 fits perfectly the data of Zhou et al.
[37], which are the most numerous among those presented.
Moreover, the averaged circular velocities of
red giants [37] have very small random errors, which
provides their large weights when searching for a solution.
In the rotation curve obtained by these authors,
there is a peak of velocities in the central region of the
Galaxy, so the potential they derived is not entirely
convenient for using in the analysis of objects located
in the central region of the Galaxy.

As can be seen from Figs. 2 and 3, at distances
$R>R_0$, there is good agreement with the linear
dependence obtained by Mroz et al. [35] for Cepheids
and with the curve of model 2-1. However, the
dependence of Mroz et al. [35] is not suitable to derive the
residual rotation velocities of stars in the inner region
of the Galaxy. Thus, in both the inner and outer
regions of the Galaxy, wherever there are Cepheids,
for which the distance estimates and kinematic data
are available, the rotation curve of our model 2-1 is
best suited for this purpose.

 {\begin{table}[t]                                    
 \caption[]
 {\small\baselineskip=1.0ex
 The values of the masses $M_d+M_h$ contained
within a sphere of radius $R$ according to model 2-1 and the
values of $M_b$, calculated as differences $(1.84-M_d-M_h)\times 10^{10}M_\odot$
and given in the $M_{\rm gal}$ units ($M_{\rm gal}=2.325\times 10^7 M_\odot$)
 }
 \label{t:2222}
 \begin{center}\begin{tabular}{|c|c|c|c|c|}\hline
  $R$~(kpc)  &
  $(M_d+M_h)\times 10^{10}M_\odot$  & $M_b\times 10^{10}M_\odot$  & $M_b$~($M_{\rm gal}$) \\\hline
             2.4  &     1.96      &   $-$0.08~~~   & $-$34~~~ \\
             2.2  &     1.67      &        0.21    &     90 \\
             2.0  &     1.41      &        0.47    &   202 \\
             1.8  &     1.15      &        0.73    &   314 \\\hline
\end{tabular}\end{center}\end{table}}
\begin{figure}[t]
{ \begin{center}
  \includegraphics[width=0.98\textwidth]{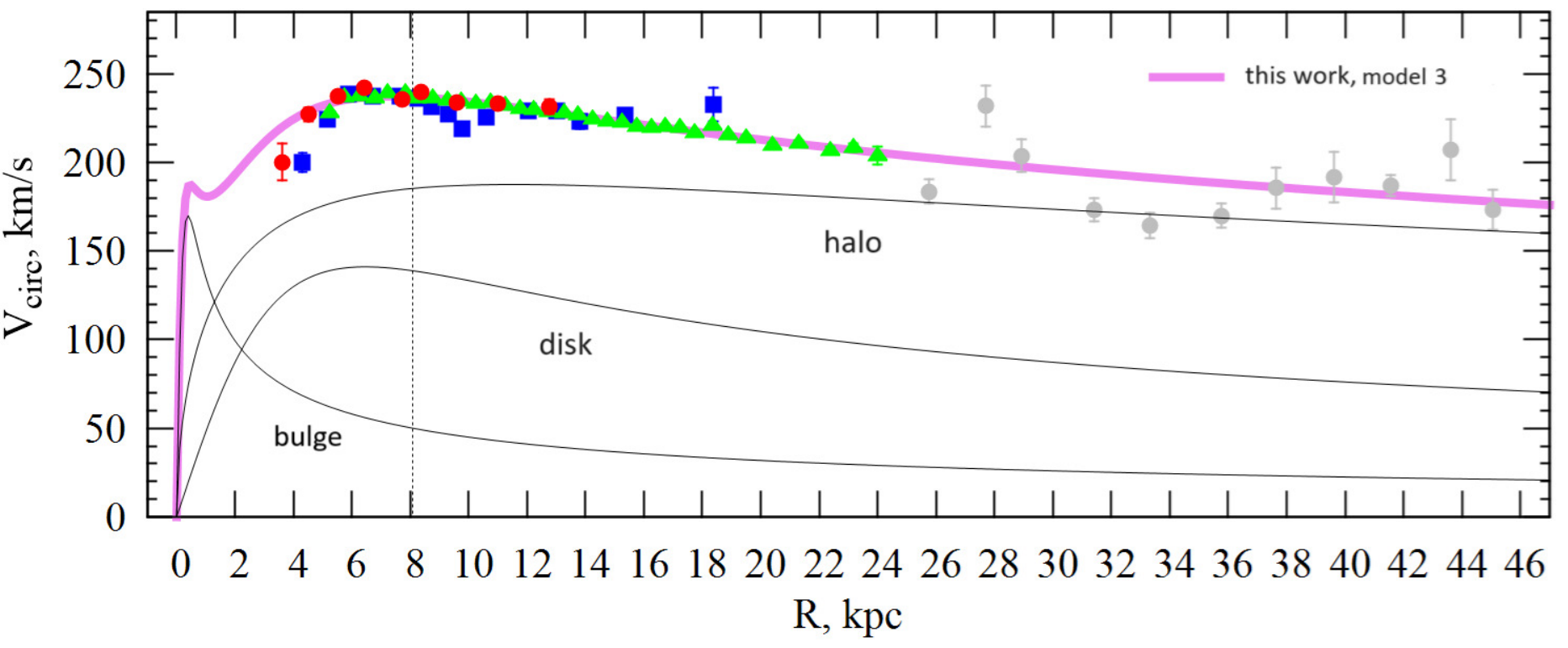}
\caption{The rotation curve of the Galaxy for model 3 (thick orange line), the vertical line marks the position of the Sun, the thin
black lines indicate the contributions of a bulge, a disk, and a halo; the velocities of classical Cepheids, masers with measured
trigonometric parallaxes, and high-luminosity red giants [37] are shown with blue squares, red circles, and green triangles,
respectively, while the gray circles show the velocities according to [1].}
 \label{f4}
\end{center}}
\end{figure}
\begin{figure}[t]
{ \begin{center}
  \includegraphics[width=0.99\textwidth]{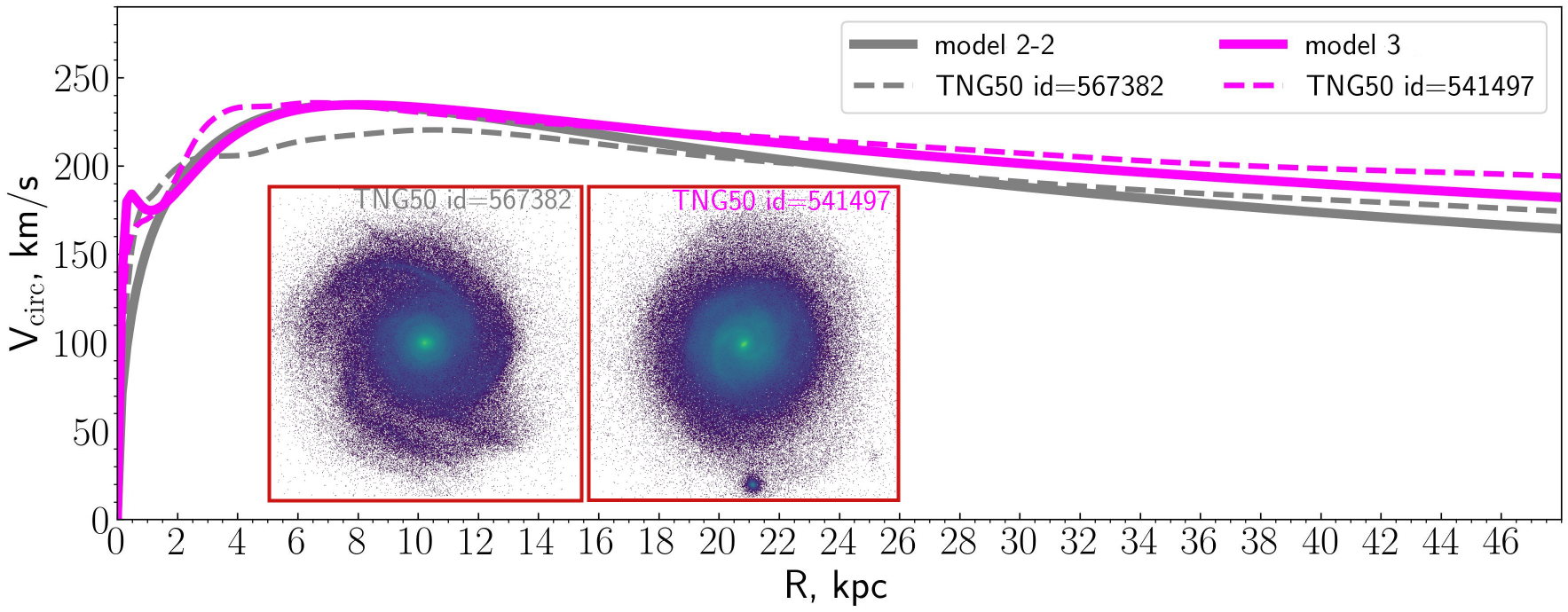}
\caption{The rotation curves according to models 2-2 and 3 (solid lines) compared to their counterparts from the TNG50 cosmological
simulations (dashed lines). The inset shows the face-on images of model galaxies according to simulations in the region
from $-$15 kpc to 15 kpc relative to the disk center.
}
 \label{nbody}
\end{center}}
\end{figure}

\subsection{Model 3}
Finally, we constructed a three-component model
that includes a bulge, a disk, and a halo. Like model 2-1,
it was built with all types of velocities considered.
Since there are no data for the central region, we set
the bulge potential manually.

The authors of [54] proposed a dynamic model of
a peanut-shaped bulge constructed with the use of
high-precision data on the Red Clump giants. The
bulge was considered as an ellipsoid with axes measuring
($2.2\times 1.4\times 1.2$)~kpc. The mass of this ellipsoid was
$1.84\times 10^{10}M_\odot$. This estimate contains the mass of the
disk $M_d$, the halo $M_h$, and the bulge itself $M_b$ within a
given volume.

With the two-component model 2-1, we calculated
the sum of the masses $M_d+M_h$, enclosed within a
sphere of radius $R$, for four values of $R$. We also determined
as a difference $(1.84-M_d-M_h)\times 10^{10}M_\odot$.
The results are presented in Table 3, and its last column
contains the bulge mass $M_b$~($M_{\rm gal}$) expressed in
units of the models we use. Note that, when fitting the
model parameters of the potential with a bar, the
potential of the bar is averaged in the form of a threeaxis
ellipsoid along the azimuth, i.e., along the angle
in the expression for the bar potential. This provides a
matching of the rotation curve with the data. However,
we plan to implement an approach that includes a bar
in the future and limit ourselves here to an axisymmetric
model of the galactic potential. So, our model 3 is
only a first approximation to reality. This concerns, in
particular, the central potential-induced burst in the
rotation curve [50].

Based on the data acquired, we adopted the following
values of the bulge parameters: $M_b=203 M_{\rm gal}$ and
$b_b=0.27$~ kpc (Table 2). Note that earlier, in our
model III [9], we used the bulge parameters with
approximately twice the mass: $M_b=443 M_{\rm gal}$ and
$b_b=0.27$~kpc. This yielded a maximum of about
260~km/s on the corresponding curve for the circular
velocity.

As a result, we constructed model 3; and the values
of its parameters are given in the last column of
Table~2. When fitting this model to the data, the
following values were obtained for the two target
parameters: $K_{z=1.1}/2\pi G=72.2~M_\odot$~pc$^{-2}$ and
$\rho_0=0.10 M_\odot$~pc$^{-3}$. Here the value of the circular rotation
velocity of the Galaxy at the near-solar distance is $V_0=234$~km/s. In addition, for model 3,
$M(<200~\hbox {kpc})=(0.60\pm0.09)\times10^{12}M_\odot,$
$M(<100~\hbox {kpc})=(0.47\pm0.10)\times10^{12}M_\odot,$
$M(<50~\hbox {kpc})=(0.35\pm0.08)\times10^{12}M_\odot$ and
$M(<30~\hbox {kpc})=(0.27\pm0.06)\times10^{12}M_\odot.$

We can compare these estimates to the limits
imposed on the mass of the Galaxy, for example, in
papers [55] and [56]: $M(<100~\hbox {kpc})=(0.56\pm0.4)\times10^{12}M_\odot$ and
$M(<32.4~\hbox {kpc})=(0.285\pm0.010)\times10^{12}M_\odot$,
respectively. Evidently, there is good agreement
between the mass estimates obtained with model
3 and in the papers mentioned.

Note that the velocity curve with a maximum of
$V_{circ}\sim150$~km/s, corresponding to the bulge contribution
to the rotation curve in model 3 (Fig. 4) with
the bulge parameter values adopted here (Table 2),
agrees well with analogous curves used by other
authors [57-59].

\subsection{Comparison to Numerical Calculations
of the Evolution of Galaxies}

By the example of the constructed models 2-2 and
3, we verified models of this type for self-consistency.
The thing is that, from a physical point of view, the
individual components must be somehow connected
to each other, since the dark halo, the stellar disk, and
the bulge should all participate in mutual gravitational
interactions. Consequently, both the parameters and,
hence, the profiles of individual components must be
implicitly interrelated in some way. It is a challenge to
determine the exact relationship here, since, in general,
a multicomponent model contains a large number
of parameters. However, if the model is physical, it
is to be formed in the course of self-consistent evolution
of galaxies, and, consequently, should also be
derived from cosmological calculations that represent
the evolution of galaxies in the Universe. Then, if we
manage to find by cosmological calculations a model
galaxy with a rotation curve close to the observed one,
we can say that such a potential could have arisen in a
self-consistently evolving galaxy. If, however, the calculations
fail to result in a model galaxy with a rotation
curve similar to that obtained, the situation is more
complicated. The absence of a model galaxy with a
rotation curve close to the observed one may suggest
that the sample of model galaxies is simply not representative.

The following experiment was carried out. Currently,
several catalogs of model galaxies similar to the
Milky Way are available. These catalogues are created
on the basis of numerical calculations performed in
the course of large projects, such as Illustris TNG50
[60], in which the physics of interaction of various
components, including gas, star formation, magnetohydrodynamics,
etc., is taken into account. We used
the catalogue of Pilepich et al. [61] constructed on the
basis of numerical calculations of this kind and containing
198 model galaxies. Next, for each model galaxy,
the rotation curve was determined, then it was
compared to the rotation curve of models 2-2 and 3,
and the standard deviation between the curves was
derived on the final grid with a step of 100 pc from 0 to
46 kpc. The rotation curves of model galaxies were
constructed by means of the AGAMA software package
[62], which was used to determine an axisymmetric
approximation of the potential and then, on its
basis, the circular velocities of stars. This program
allows one to approximate the potential of model galaxies
using cylindrical splines, which is what we did.
For both models 2-2 and 3, we found models from the
TNG50 that are closest to them in terms of the
rotation curve (see Fig.~5). In Fig.~5, the inset shows
the face-on images of model galaxies (the region from
$-$15 to $+$15 kpc relative to the center of the model).

For model 2-2, the closest rotation curve from the
TNG50, although being close to it in the central
region ($R<4$~kpc) and in the region far from the center
($R>20$~ kpc), still very poorly coincides with it in
the solar neighborhood. The values of circular velocities
obtained in the numerical calculation are significantly
lower ($\sim$200 km/s) in this area. For model 3, the
situation with the comparison turns out to be significantly
better. The rotation curve produced by numerical
calculation coincides well with that of model 3 in
the region from 6 to 24 kpc (including the vicinity of
the Sun).

The fact that, for model 3, a similar model of the
Galaxy can be found by numerical simulations may
indicate that such a model is indeed self-consistent. In
the case of model 2-2, the presence of significant discrepancies
between its rotation curve and the numerically
calculated curve in the solar neighborhood may
be caused both by the fact that the catalog itself [61]
does not cover the entire space of possible parameters
of galaxies and by the fact that model 2-2 is less physical
than model 3.

 \section{CONCLUSIONS}

In this paper the parameters of the model of the
axisymmetric gravitational potential of the Galaxy
have been refined. When constructing the rotation
curve of the Galaxy, we relied on data from the literature
concerning the velocities of masers with measured
trigonometric parallaxes, classical Cepheids, Red
Clump giants, high-luminosity giants, Blue Horizontal
Branch giants, distant halo stars, globular clusters,
and dwarf satellite galaxies of the Milky Way.

The bulge potential is presented in the form of the
Plummer potential [49], the disk potential is in the
form [50], and for the halo potential the expression
from Navarro et al. [51] is used. We constructed two
two-component models (2-1 and 2-2) of the potential
of the Galaxy, which include the contributions from a
disk and a halo of invisible matter. In model 2-1 the
original rotation curve is constructed from the data
covering a distance range of $R:3-190$~kpc, and
model 2-2 uses a shorter interval of $R:3-24$~kpc. The
three-component model 3 includes the contributions
from a disk and a halo, as well as a low-mass bulge.

The constructed models 3 and 2-2 were tested for
self-consistency by comparing their rotation curves
with a set of model ones. The rotation curves of model
galaxies were generated with the Illustris TNG50 software
package. It was concluded that model 3 is close to
self-consistent, while model 2-2 is less physical.

\section*{ACKNOWLEDGMENTS}
The authors are grateful to the reviewer for useful comments
that contributed to improving the paper.

\bigskip\bigskip{\bf REFERENCES}
\medskip {\small

1. P. Bhattacharjee, S. Chaudhury, and S. Kundu,  Astrophys. J. {\bf 785}, 63 (2014).

2. Y. Huang, X.-W. Liu, H.-B. Yuan, et al., MNRAS {\bf 463}, 2623 (2016).

3. A.T. Bajkova, V.V. Bobylev, Open Astron. {\bf 26}, 72 (2017). 

4. I. Ablimit, G. Zhao, C. Flynn, and S.A. Bird, Astrophys. J. {\bf 895}, L12 (2020).

5. Y. Jiao, F. Hammer, H. Wang, et al., Astron. Astrophys. {\bf 678}, 208 (2023).

6. G.H. Hunter, M.C. Sormani, J.P. Beckmann, et al.,  Astron. Astrophys. {\bf 692}, A216 (2024).

7. D. Erkal, V.A. Belokurov, and D.L. Parkin, MNRAS {\bf 498}, 5574, (2020).

8. A. Kravtsov, and S. Winney, O.J. Astrophys.  {\bf 7}, id. 50 (2024).

9. A.T. Bajkova, V.V. Bobylev, Astron. Lett. {\bf 42}, 567 (2016). 

10. A.T. Bajkova, V.V. Bobylev, A.O. Gromov,  Astron. Lett. {\bf 43}, 241 (2017). 

11. W.B. Burton and M.A. Gordon, Astron. Astrophys. {\bf 63}, 7 (1978).

12. D.P. Clemens, Astroph. J. {\bf 295}, 422 (1985).

13. L. Chemin, F. Renaud, and C. Soubiran, Astron. Astrophys. {\bf 578}, 14 (2015).

14. P.J. McMillan, MNRAS {\bf 465}, 76 (2017).

15. E. Vasiliev, MNRAS {\bf 482}, 1525 (2019).

16. X. Ou, A.-C. Eilers, L. Necib, and A. Frebel, MNRAS {\bf 528}, 693 (2024).

17. M.J. Reid, K.M. Menten, A. Brunthaler, et al., Astrophys. J. {\bf 885}, 131 (2019).

18. Y. Xu, S.B. Bian, M.J. Reid, et al., Astrophys. J. Suppl. {\bf 253}, 1  (2021).

19. S.B. Bian, Y. Xu, J.J. L, et al., Astron. J. {\bf 163}, 54  (2022).

20. X. Mai, B. Zhang, M.J. Reid, et al.,  Astrophys. J. {\bf 949}, 10  (2023).

21. G.N. Ortiz-Le\'on, S.A. Dzib, L. Loinard, et al., Astron. Astrophys. {\bf 673}, 1 (2023).

22. J. Ord\'o\~nez-Toro, S. A. Dzib, L.T. Loinard, et al., Astron. J. {\bf 167}, id. 108 (2024).

23. D.M. Skowron, J. Skowron, P. Mr\'oz, et al., Science {\bf 365}, 478 (2019).

24. A. Udalski, M.K. Szyma\'nski, and G. Szyma\'nski, Acta Astron. {\bf 65}, 1 (2015). 

25. I. Soszy\'nski, A. Udalski, M.K. Szyma\'nski, et al., Acta Astron. {\bf 70}, 101 (2020).

26. P. Pietrukowicz,  I. Soszy\'nski, A. Udalski,  Acta Astron. {\bf 71}, 205  (2021).

27. D.M. Skowron, R. Drimmel, S. Khanna, A. Spagna, E. Poggio, and P. Ramos,
Astrophys. J. Suppl. Ser. {\bf 278}, Issue 2, id 57 (2025).

28. Gaia Collab. (A. Vallenari, A.G.A. Brown, T. Prusti, et al.), Astron. Astrophys. {\bf 674}, 1 (2023). 

29. V.V. Bobylev, A.T. Bajkova, Research in Astron. and Astrophys. {\bf 23}, No. 4, id. 045001 (2023).

30. Gaia Collab. (A.G.A. Brown, A. Vallenari, T. Prusti,  et al.), Astron. Astrophys. {\bf 649}, 1 (2021). 

31. P. Mr\'oz, A. Udalski, D.M. Skowron, et al., Astrophys. J. {\bf 870}, L10 (2019).

32. Gaia Collab. (A.G.A. Brown, A. Vallenari, T. Prusti,  et al.), Astron. Astrophys. {\bf 616}, 1 (2018). 

33. Y. Zhou, Xi Li, Y. Huang, and H. Zhang,  Astrophys. J. {\bf 946}, 73 (2023).

34. S.R. Majewski, R.P. Schiavon, P.M. Frinchaboy, et al.,  Astrophys. J. {\bf 154}, 94 (2017).

35. A. Luo, H.T. Zhang, Y.H. Zhao, et al.,  Research in Astron. and Astrophys. {\bf 12}, 1243 (2012). 

36. L.-C.  Deng,  H.J. Newberg, C. Liu, et al.,  Research in Astron. and Astrophys. {\bf 12}, 735 (2012). 

37. J. Binney, S. Tremaine, {\it Galactic Dynamics: Second Edition},
    Published by Princeton University Press, Princeton, NJ USA,  (2008).

38. S. P\~oder, M. Benito, J. Pata, et al., 
 Astron. Astrophys. {\bf 676}, A134  (2023).

39. O. Koop, T. Antoja, A. Helmi, T.M. Callingham, and C.F. P. Laporte,  Astron. Astrophys. {\bf 692}, A50 (2024).

40. A.-C. Eilers, D.W. Hogg, H.-W. Rix, and M.K. Ness, Astrophys. J. {\bf 871}, 120 (2019).

41. V.V. Bobylev, A.T. Bajkova, Astron. Rep. {\bf 65}, 498 (2021). 

42. M. Miyamoto and R. Nagai, Publ. Astron. Soc. Japan {\bf 27}, 533 (1975).

43. J.F. Navarro, C.S. Frenk, and S.D.M. White,  Astrophys. J. {\bf 490}, 493 (1997).

44. A. Irrgang, B. Wilcox, E. Tucker, and L. Schiefelbein, Astron. Astrophys. {\bf 549}, 137 (2013).

45. J. Holmberg and C. Flynn, MNRAS {\bf 352}, 440 (2004).

46. D. Horta, A.M. Price-Whelan, D.W. Hogg, et al., Astrophys. J. {\bf 962}, 165 (2024).

47. V.V. Bobylev, A.T. Bajkova, Astron. Rep. {\bf 67}, 812 (2023b).

48. E. Vasiliev, V. Belokurov, and D. Erkal, MNRAS {\bf 501}, 2279 (2021).

49. S. Koposov, D. Erkal, T.S. Li, et al., MNRAS {\bf 521}, 4936 (2023).

50. A. Pillepich, D. Nelson, V. Springel, et al., MNRAS {\bf 490}, 3196 (2019).

51. A. Pillepich, D. Sotillo-Ramos, R. Ramesh, et al., MNRAS {\bf 535}, 1721 (2024).

52. E. Vasiliev, MNRAS {\bf 484}, 2832 (2019).

53. R. Sch\"onrich, J. Binney, and W. Dehnen , MNRAS {\bf 403}, 1829 (2010).

54. M. Portail, C. Wegg, O. Gerhard, and I. Martinez-Valpuesta, MNRAS {\bf 448},  713 (2015). 

55. P.F. de Salas, K. Malhan, K. Freese, K. Hattori, and M. Valluri,
 Journal of Cosmology and Astroparticle Physics  {\bf 10}, id. 037 (2019).

56. X. Ou , A.-C. Eilers, L. Necib, and A, Frebel, MNRAS {\bf 528},  693  (2024).

57. G.H. Hunter, M.C. Sormani, J.P. Beckmann, E. Vasiliev, S.C.O. Glover, R.S. Klessen, J.D. Soler, N. Brucy,
 et al., Astron. Astrophys. {\bf 692}, A216 (2024).

58. H.C. Plummer, MNRAS {\bf 71}, 460 (1911).

59.  S. K. Kataria, and M. Das, MNRAS {\bf 475}, 1653 (2018).

60.S. Khrapov, A. Khoperskov, and V. Korchagin, Galaxies {\bf 9}, id. 29 (2021).

61. M. Portail, O. Gerhard, C. Wegg, and M. Ness, MNRAS {\bf 465}, 1621 (2017).

62. Z. Li, J. Shen, O. Gerhard, and J.P. Clarke, Astrophys. J. {\bf 925}, 71 (2022).
}
\end{document}